\documentclass[12pt]{article}
\usepackage{amsmath}
\usepackage{graphicx}
\usepackage{enumerate}
\usepackage{natbib}
\usepackage{url} % not crucial - just used below for the URL 
\usepackage{enumitem}

\usepackage{econometrics} % added by me
\usepackage{amssymb} % added by me
\usepackage{bm} % added by me
\usepackage{bbm} % added by me
\usepackage{cases} % added by me
\usepackage{color} % added by me
\usepackage{comment} % added by me
\usepackage{hyperref}
\usepackage{caption} % added by Simon
\usepackage{subcaption} % added by Simon
\newcommand{\blind}{1}

\begin{document}

\def\spacingset#1{\renewcommand{\baselinestretch}%
{#1}\small\normalsize} \spacingset{1}

%%%%%%%%%%%%%%%%%%%%%%%%%%%%%%%%%%%%%%%%%%%%%%%%%%%%%%%%%%%%%%%%%%%%%%%%%%%%%%

%\newpage
\spacingset{1.1} % DON'T change the spacing!

\if1\blind
{
  \title{\bf Short $\&$ simple introduction to
  \\ Bellman filtering and smoothing}
  \author{Rutger-Jan Lange\thanks{I thank Dick van Dijk, Simon Donker van Heel and Ramon de Punder for useful comments.}\\
    Econometric Institute, Erasmus University Rotterdam, Netherlands}
  \maketitle
} \fi

\if0\blind
{
  \bigskip
  \bigskip
  \bigskip
  \begin{center}
    {\LARGE\bf Title}
\end{center}
  \medskip
} \fi

\bigskip

\begin{abstract}
    \noindent Based on \citeauthor{bellman1957dynamic}'s dynamic-programming principle, \cite{lange2024bellman} presents an approximate method for filtering, smoothing and parameter estimation for possibly non-linear and/or non-Gaussian state-space models. While the approach applies more generally, this pedagogical note highlights the main results in the case where (i) the state transition remains linear and Gaussian while (ii) the observation density is log-concave and sufficiently smooth in the state variable. I
    demonstrate how \citeauthor{kalman1960new}'s (\citeyear{kalman1960new}) filter and \citeauthor{rauch1965maximum}'s (\citeyear{rauch1965maximum}) smoother can be obtained as special cases within the proposed framework. The main aim 
    %of this short note 
    is to present non-experts (and my own students) with an accessible introduction, enabling them to implement the proposed methods.
    %for filtering and smoothing.
    %I mostly follow the notation for state-space models as in \cite{harvey1990forecasting} and \cite{durbin2012time}.
\end{abstract}

\noindent%
{\it Keywords:}  state-space model, Kalman filter, Rauch Tung Striebel smoother, particle filter
% \color{blue} 3 to 6 keywords, that do not appear in the title\color{black}
%\vfill

\section{State-space model}

\textbf{State-space model with general observation density.} For discrete times $t=1,\ldots,n$, a state-space model with a linear Gaussian state-transition equation but a general observation density can be written as
\begin{align}
\label{observation equation}
\text{observation equation:}&&\quad \bm{y}_t &\sim p(\bm{y}_t|\bm{x}_t),
\\
\label{state equation}
\text{state-transition equation:}&&\quad \bm{x}_{t} &= \bm{c}\,+\, \bm{T}\,\bm{x}_{t-1} \,+ \,\bm{R}\,\bm{\eta}_t,\quad \bm{\eta}_t \sim \text{i.i.d.}\mathrm{N}(\bm{0},\bm{Q}),
\end{align}
which roughly matches the notation in \cite{harvey1990forecasting} and \citet[p.\ 209-10]{durbin2012time}. These authors write $\bm{\alpha}_t$ where I write $\bm{x}_t$; the latter is apparently more common in statistics and engineering (e.g.\ \citealp{katzfuss2016understanding}). It is implicitly assumed that $\bm{x}_t$ takes values in $\mathbb{R}^d$, where $d$ is the dimension of the state space. The dimension of the column vector $\bm{y}_t$ is left implicit and need \emph{not} match that of $\bm{x}_t$.

\textbf{Observation equation.} Equation~\eqref{observation equation} says that $\bm{y}_t$ is drawn from a conditional density $p(\cdot|\bm{x}_t)$. While purists may object to the (abuse of) notation, equation~\eqref{observation equation} is nevertheless common in econometrics (e.g.\ \citealp[pp.\ 209-10]{durbin2012time}).
 The observation density $p(\cdot|\bm{x}_t)$ may further depend on static shape parameters or exogenous variables that are known at time $t-1$, but for simplicity these dependencies are suppressed. When the data $\bm{y}_t$ are discrete (e.g.\ in the case of count data), $p(\cdot|\bm{x}_t)$ can be interpreted as a probability rather than a density. 
 %I assume that $\log p(\bm{y}|\bm{x})$ is well defined for all relevant values of $\bm{y}$ and $\bm{x}$. 
 For each (fixed) $\bm{y}$, our standing assumptions are that $\log p(\bm{y}|\bm{x})$ is (i) concave and (ii) twice continuously differentiable in the state variable $\bm{x}\in \mathbb{R}^d$. These assumptions imply that the Hessian matrix of $\log p(\bm{y}|\bm{x})$ with respect to $\bm{x}$ is well-defined and negative semi-definite for all $\bm{x}$ and all $\bm{y}$.

\textbf{State-transition equation.} Equation~\eqref{state equation} says that the unobserved state $\bm{x}_t$ follows  linear dynamics with an intercept $\bm{c}$, transfer matrix $\bm{T}$, selection matrix $\bm{R}$ (which typically contains ones and zeros) and independent and identically distributed (i.i.d.)~Gaussian increments $\bm{\eta}_t$ with mean zero and positive semi-definite covariance matrix~$\bm{Q}$, where the nomenclature is taken from \citet[pp.~43-4]{durbin2012time}.  It is convenient (and nonrestrictive) to assume that the transfer matrix $\bm{T}$ contains at least \emph{some} non-zero elements; otherwise, there would be no persistence in the latent state. 
%It is also implicitly assumed that the increments $\bm{R}\bm{\eta}_t$ have a finite second moment, i.e.\ $\bm{R}\bm{Q}\bm{R}'$, where the prime denotes a transpose, contains only finite elements. 
For simplicity I assume that $\bm{c}$, $\bm{T}$, $\bm{R}$ and $\bm{Q}$ are static, but following \citet[p.~43]{durbin2012time} these quantities could be allowed to depend deterministically or stochastically on time; at least, as long as the relevant quantities at time $t$ are known to the researcher at time $t-1$.

\textbf{Three aims.} We are interested in (i) filtering, (ii) smoothing and (iii) parameter estimation. 
%The aim of filtering and smoothing is to estimate the states given the observations. 
Filtering means estimating the most \emph{recent} state $\bm{x}_t$ for all $t\leq n$ based on the real-time (i.e.\ expanding) information set $\{\bm{y}_i\}_{i=1,\ldots,t}$. Smoothing means estimating \emph{all} states $\{\bm{x}_i\}_{i=1,\ldots,n}$ based on the whole (i.e.\ fixed) data set $\{\bm{y}_i\}_{i=1,\ldots,n}$. The third aim is to estimate, additionally, the static parameters in equations~\eqref{observation equation}--\eqref{state equation}.
%The parameter-estimation problem considers both the unobserved states $\{\bm{x}_t\}_{t}$ \emph{and} the static parameters to be unknown. 

\textbf{Special case: Linear Gaussian observation equation.} A famous example of the general state-space model above is the linear Gaussian state-space model, in which case the general observation equation~\eqref{observation equation} is replaced by its linear Gaussian version that in the notation of \citet[pp.~100-1]{harvey1990forecasting} or \citet[p.~85]{durbin2012time}
reads
\begin{equation}
\label{observation equation Kalman filter}
\text{observation equation:}\qquad 
 \bm{y}_t =\bm{d}\,+\, \bm{Z}\,\bm{x}_t\,+\,\bm{\varepsilon}_t,\qquad \bm{\varepsilon}_t \sim \text{i.i.d.} \mathrm{N}(\bm{0},\bm{H}),
\end{equation}
while the state transition follows equation~\eqref{state equation}. As the entire state-space model~\eqref{state equation}--\eqref{observation equation Kalman filter} is now linear and Gaussian, this is the standard case to which \citeauthor{kalman1960new}'s (\citeyear{kalman1960new}) filter and \citeauthor{rauch1965maximum}'s (\citeyear{rauch1965maximum}) smoother apply. Here, $\bm{y}_t$ is a linear transformation of the latent state $\bm{x}_t$, with intercept $\bm{d}$ and coefficient matrix $\bm{Z}$, while the additive observation noise $\bm{\varepsilon}_t$ is i.i.d.\ Gaussian with mean zero and covariance matrix $\bm{H}$, which for simplicity is assumed to be positive definite. (By a standard limiting argument, the results below remain valid if $\bm{H}$ is only positive \emph{semi}-definite.) It is again convenient (and nonrestrictive) to assume that $\bm{Z}$ contains at least \emph{some} non-zero elements, as otherwise the observations would be entirely uninformative about states.

\section{Statement of Bellman filter for model~\eqref{observation equation}--\eqref{state equation}}

\textbf{Background.} Using \citeauthor{bellman1957dynamic}'s (\citeyear{bellman1957dynamic}) dynamic-programming principle, \cite{lange2024bellman} derives an approximate filter based on recursively estimating the most likely value of $\bm{x}_t$, i.e.\ the mode.
%The solution to Bellman's equation is known as the \emph{value function} and its argmax, at each time step, serves as our filtered state. 
%While Bellman's equation cannot typically be solved exactly, its solution can be approximated, at each time step, by a multivariate quadratic function.
%; this can be viewed as a form of (parametric) approximate dynamic programming. 
%While non-parametric approaches may be possible, these have not yet been explored. 
%As it turns out, this quadratic approximation is exact for the linear Gaussian state-space model~\eqref{state equation}--\eqref{observation equation Kalman filter}, yielding \citeauthor{kalman1960new}'s (\citeyear{kalman1960new}) filter as a special case of the Bellman filter, as explained below in section~\ref{sec3}. 
While the approach is generally applicable, the resulting filter takes a particularly attractive form if we are content to maintain the classic assumption of a linear Gaussian state equation~\eqref{state equation}, while allowing the general observation density~\eqref{observation equation}.
%In this note, therefore, we highlight the results for the (often sufficiently general) state-space model~\eqref{observation equation}--\eqref{state equation}, which contains the classic model~\eqref{state equation}--\eqref{observation equation Kalman filter} as a special case.
%(see section~\ref{sec3}).

\textbf{Notation.}
Like the Kalman filter, the Bellman filter keeps track of one-step-ahead predicted estimates of $\bm{x}_t$, denoted $\hat{\bm{x}}_{t|t-1}$, and real-time (i.e.\ `filtered') estimates of $\bm{x}_t$, denoted $\hat{\bm{x}}_{t|t}$.
%The hats indicate that these quantities represent the researcher's estimates, while the subscript indicates the state that is estimated conditional on the relevant information set. 
The corresponding measures of uncertainty are denoted $\bm{P}_{t|t-1}$ and $\bm{P}_{t|t}$, respectively, which can be roughly interpreted as covariance matrices; this interpretation is exact when the Kalman filter applies.
%; this interpretation is exact when the Kalman filter applies.
%for the linear Gaussian observation equation~\eqref{observation equation Kalman filter}.
%The difference with the Kalman filter is that the Bellman filter is more generally applicable; however, the price paid for this generality is that it is typically only approximate (i.e., unless it equals the Kalman filter).  

\textbf{Fisher's information.} 
%If $\log p(\bm{y}_t|\bm{x}_t)$ is twice continuously differentiable in $\bm{x}_t$, with  probability one in $\bm{y}_t$, as I assume throughout, then 
Fisher's information matrix (in $\mathbb{R}^{d\times d}$) is defined as an expectation, over $\bm{y}$, of the negative Hessian matrix of $\log p(\bm{y}|\bm{x})$, conditional on $\bm{x}$, i.e.\
\begin{equation}
\label{Fisher}
\bm{\mathcal{I}}(\bm{x}) \; :=\; 
%\mathrm{E}\left[- \nabla^2 \log p(\bm{y}_t | \bm{x}_t) \right] = 
\int \left[- \nabla^2 \log p(\bm{y} | \bm{x}) \right]p(\bm{y}|\bm{x})\,\mathrm{d}\bm{y},\qquad \bm{x}\in \mathbb{R}^d,
\end{equation}
where the integral can be replaced by a summation in the case of discrete observations. The information quantity $\bm{\mathcal{I}}(\bm{x})$ can be evaluated at any state variable $\bm{x}$; below, we need $\bm{\mathcal{I}}(\hat{\bm{x}}_{t|t})$. Throughout, the gradient $\nabla:=\mathrm{d}/\mathrm{d}\bm{x}$ and Hessian operator $\nabla^2:=\nabla \nabla' = \mathrm{d}^2/(\mathrm{d}\bm{x}\mathrm{d}\bm{x}')$, where a prime denotes a transpose, are understood as acting on the state variable $\bm{x}$. Fisher's information $\bm{\mathcal{I}}(\bm{x})$ can often be computed in closed form; moreover, it is positive semi-definite under our standing assumptions.
%^which imply that its negative Hessian matrix is positive semi-definite. 
%imply that the negative Hessian matrix is positive semi-definite.
%under common identification assumptions, which I assume are satisfied.

\textbf{Prediction step.} Let $t\geq 1$ and suppose that $\bm{P}_{t-1|t-1}$ is positive definite with bounded eigenvalues. The proposed prediction steps in \citet[Table 3]{lange2024bellman} coincide with those of the Kalman filter (e.g. \citealp[pp.~105-6]{harvey1990forecasting} and \citealp[p.~86]{durbin2012time}):
\begin{align}
    \hat{\bm{x}}_{t|t-1}&=\bm{c}+\bm{T}\,\hat{\bm{x}}_{t-1|t-1},
    \label{level prediction}
    \\
    \bm{P}_{t|t-1} &=\bm{T} \bm{P}_{t-1|t-1} \bm{T}'+\bm{R}\,\bm{Q}\,\bm{R}'.
    \label{uncertainty prediction}
\end{align}
Intuitively, the level prediction~\eqref{level prediction} mimics the true dynamics~\eqref{state equation}, while the uncertainty prediction~\eqref{uncertainty prediction} can be analogously derived. The predicted uncertainty measure $\bm{P}_{t|t-1}$ is positive definite, because  $\bm{P}_{t-1|t-1}$ is assumed positive definite and $\bm{T}$ contains at least some non-zero elements, while $\bm{R}\bm{Q}\bm{R}'$ is positive semi-definite (as $\bm{Q}$ is positive semi-definite). 

\textbf{Updating step.} The Bellman-filtered update as postulated in \citet[eq.\ 16]{lange2024bellman} is
\begin{align}
\hat{\bm{x}}_{t|t}&=\arg \max_{\bm{x}\in \mathbb{R}^d} \Big\{ \log p(\bm{y}_t|\bm{x})-\frac{1}{2}(\bm{x}-\hat{\bm{x}}_{t|t-1})'\bm{P}_{t|t-1}^{-1}(\bm{x}-\hat{\bm{x}}_{t|t-1})\Big\},
\label{level update}
\\
\bm{P}_{t|t}&= \big[\bm{P}_{t|t-1}^{-1} +\bm{\mathcal{I}}(\hat{\bm{x}}_{t|t})\big]^{-1} ,
\label{uncertainty update}
\end{align}
where $(\cdot)^{-1}$ denotes a matrix inverse. 

\textbf{Interpretation of level update~\eqref{level update}.} Equation~\eqref{level update} says that the researcher's filtered state $\hat{\bm{x}}_{t|t}$ maximises the current log-likelihood contribution, $\log p(\bm{y}_t|\bm{x})$, subject to a quadratic penalty centred around the prediction $\hat{\bm{x}}_{t|t-1}$. The strength of the penalty is determined by $\bm{P}_{t|t-1}^{-1}$, such that the penalty is weaker when the predicted uncertainty is larger; in effect, more weight is then placed on the observation $\bm{y}_t$, while the influence of the prediction $\hat{\bm{x}}_{t|t-1}$ is diminished. This interpretation reveals a close connection with stochastic proximal point methods (e.g.\ \citealp{toulis2017asymptotic}, \citealp{patrascu2018nonasymptotic} and \citealp{asi2019stochastic}), because update~\eqref{level update} can be viewed as computing a new `point' $\hat{\bm{x}}_{t|t}$ that is `stochastic' due to the observation~$\bm{y}_t$, while remaining `proximal' (i.e.\ close) to the prediction $\hat{\bm{x}}_{t|t-1}$. Hence the Kalman filter---a special case of equation~\eqref{level update}, as shown in section~\ref{sec3}---is, in fact, a stochastic proximal point method; to my knowledge, this fact has not been explicitly recognised. 
%not widely known.
%the Kalman filter (which, as shown in section~\ref{sec3}, is a special case of equation~\eqref{level update}) has not before been identified as a stochastic proximal point method.
This may be because proximal methods are typically used for gradually `learning' unknown \emph{static} parameters (see section~\ref{sec:online}), while we are interested in tracking a \emph{time-varying} unobserved state. Recasting the Kalman filter as an instance of a (dynamic) proximal method is relevant as the latter permits favourable properties that are well known in the literature (e.g.\ stability and contractivity).
%for a comparison.

Equation~\eqref{level update} can also be understood as computing the posterior mode in a Bayesian setting, in which case the quadratic penalty is interpreted as a Gaussian prior with mean $\hat{\bm{x}}_{t|t-1}$ and covariance matrix $\bm{P}_{t|t-1}$. This interpretation reveals a connection with Laplace approximations methods in both the Bayesian (e.g.\ \citealp{rue2009approximate}) and frequentist literature (e.g.\ \citealp{koyama2010approximate}). The analogy with these methods breaks down when considering more general dynamics than the linear Gaussian state equation~\eqref{state equation}; see \citet[sec.~3]{lange2024bellman} for a discussion of the Bellman filter in the more general case. 

\textbf{Interpretation of uncertainty update~\eqref{uncertainty update}.} Equation~\eqref{uncertainty update} says that the updated measure of uncertainty, $\bm{P}_{t|t}$, is the inverse of the expectation of the negative Hessian matrix of the objective function, evaluated at the peak. As Fisher's information is positive semi-definite,  the resulting matrix $\bm{P}_{t|t}$ is positive definite as desired. Several authors (e.g.\  \citealp[p. 504]{fahrmeir1992posterior} and \citealp[eq.~36]{lambert2022recursive}) write Kalman's covariance update as in our
equation~\eqref{uncertainty update}, but their generalisations of Kalman's level update differ from ours. 

In equation~\eqref{uncertainty update}, Fisher's information quantity $\bm{\mathcal{I}}(\hat{\bm{x}}_{t|t})$ may be replaced, when convenient, by its `realised' version $-\nabla^2\log p(\bm{y}_t|\hat{\bm{x}}_{t|t})$, which is similarly positive semi-definite under our standing assumptions. Indeed, this would yield the approximate covariance update in \citet[eq.~2.1] {ollivier2018online}. As suggested in \citet[sec.~8]{lange2024bellman}, we may also employ a weighted average of $\bm{\mathcal{I}}(\hat{\bm{x}}_{t|t})$ and $-\nabla^2\log p(\bm{y}_t|\hat{\bm{x}}_{t|t})$. While we have assumed that $\log p(\bm{y}_t|\bm{x})$ is concave in $\bm{x}$, taking a weighted average may be particularly useful when this assumption is violated; in this case, a sufficiently large weight should be placed on the Fisher information matrix to ensure that the weighted sum is positive definite. These approaches are identical for the linear Gaussian observation equation~\eqref{observation equation Kalman filter}, for then, as shown in section~\ref{sec3}, the negative Hessian is constant and (hence) equals the Fisher information matrix.

\textbf{Initialisation.} Transformations~\eqref{uncertainty prediction} and~\eqref{uncertainty update} preserve positive definiteness of the uncertainty measures $\smash{\{\bm{P}_{t|t-1}\}_{t=1,\ldots,n}}$ and $\smash{\{\bm{P}_{t|t}\}_{t=1,\ldots,n}}$ under our standing assumptions on $\log p(\bm{y}|\bm{x})$ along with the fact that $\bm{T}$ contains at least some non-zero elements. This means that the Bellman filter can be initialised with any positive definite matrix $\bm{P}_{0|0}$ (and an arbitrary $\hat{\bm{x}}_{0|0}$), such that recursions~\eqref{level prediction} through~\eqref{uncertainty update} remain well-defined for all $t=1,\ldots,n$. 

\subsection{Computing the filtered state}
Importantly for computation, $\hat{\bm{x}}_{t|t}$ is guaranteed to exist (i.e.\ uniquely), because the objective function, in curly brackets in equation~\eqref{level update}, is \emph{strongly concave} in $\bm{x}\in \mathbb{R}^d$. Strong concavity of a twice-continuously differentiable function means that 
%all eigenvalues of the Hessian matrix are negative on the entire state space, while 
the largest eigenvalue of the Hessian matrix is negative and bounded away from zero by some (positive) constant on the entire state space. 
Intuitively, the objective function looks like 
%This means that the function is always curving downward by at least some minimum amount, like an i
an inverted parabola, for which the second derivative is similarly negative and uniformly bounded away from zero.

In equation~\eqref{level update}, the Hessian matrix $\smash{\nabla^2\log p(\bm{y}_t|\bm{x})-\bm{P}_{t|t-1}^{-1}}$ satisfies these criteria under our standing assumptions on $\log p(\bm{y}|\bm{x})$ in combination with the positive definiteness of $\smash{\bm{P}_{t|t-1}^{-1}}$. As the Hessian of $\log p(\bm{y}_t|\bm{x})$ is only negative \emph{semi}-definite, the smallest eigenvalue of $\bm{P}_{t|t-1}^{-1}$ being positive is crucial in delivering uniformity; this is guaranteed if $\bm{P}_{t|t-1}$ has finite (positive) eigenvalues.
%, which is realistic if \emph{some} information regarding each element of $\bm{x}_t$ before observing $\bm{y}_t$; i.e. if  having finite eigenvalues implies that the smallest eigenvalue of its inverse is non-zero).
%The largest eigenvalue of $\bm{P}_{t|t-1}$ is bounded above, which is reasonable . %In sum, the quadratic penalty is crucial in making the objective function strongly concave. 
It follows that $\hat{\bm{x}}_{t|t}$ is the location of the unique stationary point, which can typically be found using standard (e.g.\ quasi-Newton) optimisation techniques. Users may implement their own optimisation algorithms or employ black-box optimisation tools. It may occasionally be useful to note that, even if $\log p(\bm{y}|\bm{x})$ fails to be (globally) concave in $\bm{x}$, optimisation~\eqref{level update} remains well defined if the smallest eigenvalue of $\bm{P}_{t|t-1}^{-1}$ is sufficiently large; e.g.\ \citet[sec.~8]{lange2024bellman} encounters no difficulties in several non-concave cases.

\subsection{Theoretical guarantees}
\label{sec22}

While the conditions discussed above guarantee that the Bellman filter can be feasibly implemented, they are not (yet) sufficient to guarantee any particular performance level. For example, we would like $\bm{\mathcal{I}}(\hat{\bm{x}}_{t|t})$ to remain positive definite in the long run, as this would ensure, via recursions~\eqref{uncertainty prediction} and \eqref{uncertainty update}, that the largest eigenvalue of $\bm{P}_{t|t}$ remains uniformly bounded over time. We can guarantee this by requiring the log-likelihood contribution $\log p(\bm{y}|\bm{x})$ \emph{itself} to be strongly concave in $\bm{x}$ (for all $\bm{y}$), as would the case for the linear Gaussian observation equation~\eqref{observation equation Kalman filter}. It can then be shown that the updating step~\eqref{level update} is contractive to a small region around the true state at every time step \citep[sec.~5]{lange2024bellman}, where the `minimal contraction strength' is the same for all time steps. Under some additional conditions, the mean squared filtering error can then be shown to remain uniformly bounded over time (\citealp[sec.~5]{lange2024bellman}).

\section{Online learning as a special case}
\label{sec:online}

In the online-learning literature (e.g.\ \citealp{orabona2019modern}), the unobserved state $\bm{x}_t$ is static (i.e.\ constant over time); hence, it is typically viewed as an unknown parameter rather than an unobserved state. The researcher is then interested in gradually `learning' this parameter as more data becomes available. This setup can be viewed as a special case of the state-transition equation~\eqref{state equation} by taking $\bm{c}=\bm{0}$ and $\bm{Q}=\bm{0}$, while setting $\bm{T}$ equal to the identity matrix. We then have a setting similar to \citet[eq.~7]{toulis2017asymptotic}, who perform the parameter update, at each time step, using the stochastic proximal method~\eqref{level update}. 

Under the above constraints on the state-transition equation (i.e.\ $\bm{c}=\bm{0}$, $\bm{Q}=\bm{0}$, and $\bm{T}$ being the identity), equations~\eqref{uncertainty prediction} and~\eqref{uncertainty update} yield  $\bm{P}_{t|t-1}=\bm{P}_{t-1|t-1}$ and
$\bm{P}_{t|t}^{-1}=\bm{P}_{t-1|t}^{-1}+\bm{\mathcal{I}}(\hat{\bm{x}}_{t|t})$, respectively. Together, this implies that $\bm{P}_{t|t}^{-1}$ is bigger, in a positive semi-definite sense, than $\bm{P}_{t-1|t-1}^{-1}$ at every time step. If the additive part $\bm{\mathcal{I}}(\hat{\bm{x}}_{t|t})$ is positive definite in the long run, then $\bm{P}_{t|t}^{-1}= O(t)$, such that $\bm{P}_{t|t}=O(1/t)$. This means that our estimate $\hat{\bm{x}}_{t|t}$ converges, in the long run, to a constant vector; i.e.\ eventually we `learn' the true state with perfect precision (e.g.\ \citealp[p.~2938]{ollivier2018online}). When $\bm{Q}$ is nonzero, however, the Bellman filter remains perpetually responsive, which makes sense when the true state continues to vary. We are then interested in `tracking' some path rather than `learning' some parameter; as such, there is no longer any `convergence' to a static long-run equilibrium.

\section{Kalman filter as a special case}
\label{sec3}

Predictions~\eqref{level prediction}--\eqref{uncertainty prediction} are identical to those in the Kalman filter. While the Bellman filter's updating steps~\eqref{level update}--\eqref{uncertainty update} are formulated more generally, they collapse to the standard expressions in the setting of the linear Gaussian observation equation~\eqref{observation equation Kalman filter}, as shown here.

\textbf{Score and Hessian.} For the linear Gaussian observation equation~\eqref{observation equation Kalman filter}, the logarithmic observation density along with the score and Hessian read
\allowdisplaybreaks
\begin{align}
\text{log density:} && \log p(\bm{y} | \bm{x}) & \propto -\smash{\frac{1}{2}} (\bm{y}-\bm{d}-\bm{Z}\bm{x})'\bm{H}^{-1}(\bm{y}-\bm{d}-\bm{Z}\bm{x}),
\notag
 \\
\text{score:}&& \nabla \log p(\bm{y}|\bm{x}) &= \bm{Z}'\bm{H}^{-1}(\bm{y}-\bm{d}-\bm{Z}\bm{x}),
\label{score}
\\
\label{Hessian}
\text{Hessian:}&&\nabla^2 \log p(\bm{y}|\bm{x}) &=- \bm{Z}'\bm{H}^{-1}\bm{Z}.
\end{align}
The score~\eqref{score} is linear in $\bm{x}$; hence, the Hessian matrix~\eqref{Hessian} is constant. Hence Fisher's information~\eqref{Fisher} equals $\bm{\mathcal{I}}(\bm{x})  = \bm{Z}'\bm{H}^{-1}\bm{Z}$, which is positive definite as $\bm{H}$ is assumed positive definite, while $\bm{Z}$ contains at least \emph{some} non-zero elements. The Hessian matrix is negative definite for the same reasons; hence, $\log p (\bm{y}|\bm{x})$ is strongly concave in $\bm{x}$ (for all $\bm{y}$) such that the theoretical guarantees in section~\ref{sec22} apply.

\textbf{Level update.} To show that the Kalman-filter level update is a special case of the Bellman filter's optimisation~\eqref{level update}, I write down the first-order condition of the latter, i.e.\ $\bm{0}=\nabla \log p(\bm{y}_t |\hat{\bm{x}}_{t|t})-\bm{P}_{t|t-1}^{-1}(\hat{\bm{x}}_{t|t}-\hat{\bm{x}}_{t|t-1})$, and substitute the score~\eqref{score} to yield
\begin{equation*}
    \label{foc2}
    \bm{0}=\bm{Z}'\bm{H}^{-1}(\bm{y}_t-\bm{d}-\bm{Z}\hat{\bm{x}}_{t|t})-\bm{P}_{t|t-1}^{-1}(\hat{\bm{x}}_{t|t}-\hat{\bm{x}}_{t|t-1}).
\end{equation*}
By several steps of straightforward linear algebra, this equation can be solved for $\hat{\bm{x}}_{t|t}$ to yield the classic Kalman-filter update: 
\begin{align}
\small
    \hat{\bm{x}}_{t|t} &\small = (\bm{Z}'\bm{H}^{-1}\bm{Z}+\bm{P}^{-1}_{t|t-1})^{-1}[\bm{Z}'\bm{H}^{-1}(\bm{y}_t-\bm{d})+\bm{P}_{t|t-1}^{-1} \hat{\bm{x}}_{t|t-1}],
    \notag
    \\
    &\small = (\bm{Z}'\bm{H}^{-1}\bm{Z}+\bm{P}^{-1}_{t|t-1})^{-1}[\bm{Z}'\bm{H}^{-1}(\bm{y}_t-\bm{d})+(\bm{P}_{t|t-1}^{-1}+\underbrace{\bm{Z}'\bm{H}^{-1}\bm{Z}-\bm{Z}'\bm{H}^{-1}\bm{Z}}_{=\bm{0}}) \hat{\bm{x}}_{t|t-1}],
        \notag
        \\
        &\small= \hat{\bm{x}}_{t|t-1}+ (\bm{Z}'\bm{H}^{-1}\bm{Z}+\bm{P}^{-1}_{t|t-1})^{-1} \bm{Z}' \bm{H}^{-1} (\bm{y}_t-\bm{d}-\bm{Z}\hat{\bm{x}}_{t|t-1}),       \notag
              \\
        &\small= \hat{\bm{x}}_{t|t-1}+ \bm{P}_{t|t-1}\bm{Z}'(\bm{Z}\bm{P}_{t|t-1}\bm{Z}'+\bm{H})^{-1} (\bm{y}_t-\bm{d}-\bm{Z}\hat{\bm{x}}_{t|t-1}),
        \label{Kalman filter level update}
     \end{align}
where the last line, which follows from a standard matrix-inversion lemma (see equation~\eqref{matrixlemma2} below), is the standard Kalman-filter level update; see e.g.\ \citet[p.~106]{harvey1990forecasting} or \citet[p.~86]{durbin2012time}.

\textbf{Uncertainty update.} Taking the Bellman-filter uncertainty update~\eqref{uncertainty update} and substituting the specific (constant) Fisher information $\bm{\mathcal{I}}(\bm{x})  = \bm{Z}'\bm{H}^{-1}\bm{Z}$ yields
\begin{align}
\bm{P}_{t|t}=[\bm{P}_{t|t-1}^{-1}+\bm{\mathcal{I}}(\hat{\bm{x}}_{t|t})]^{-1}&=[\bm{P}_{t|t-1}^{-1}+\bm{Z}'\bm{H}^{-1}\bm{Z}]^{-1},
\notag
\\
&=
\bm{P}_{t|t-1}-\bm{P}_{t|t-1}\bm{Z}' (\bm{Z}\bm{P}_{t|t-1}\bm{Z}'+\bm{H})^{-1}\bm{Z}\bm{P}_{t|t-1}.
\label{Kalman filter covariance update}
\end{align}
The second line, which employs the Woodbury matrix-inversion lemma (see equation~\eqref{matrixlemma1} below), is exactly Kalman's  covariance-matrix updating step; again, see \citet[p.~106]{harvey1990forecasting} or \citet[p.~86]{durbin2012time}.  

In the context of the linear Gaussian observation equation~\eqref{observation equation Kalman filter}, therefore, the Bellman-filter updating steps~\eqref{level update} and~\eqref{uncertainty update} reduce to the standard Kalman-filter updating steps~\eqref{Kalman filter level update} and~\eqref{Kalman filter covariance update}, respectively. Equations~\eqref{level update} and~\eqref{uncertainty update} have the possible advantage that they are easier to grasp, at an intuitive level, than the more specific results~\eqref{Kalman filter level update} and~\eqref{Kalman filter covariance update}, while also suggesting immediate extensions of the latter when the assumptions of a linear and/or Gaussian observation equation fail.  
%of these classic Kalman-filtering formulas.

\subsection{(Iterated) extended Kalman filter as a special case}

The link between optimisation methods and the Kalman filter has been known since \cite{bell1993iterated}, at least in the context of nonlinear models with additive Gaussian disturbances. To derive these classic results in our setup, we may consider the non-linear (but still Gaussian) observation equation $\bm{y}_t=\bm{d}+\bm{Z}(\bm{x}_t)+\bm{\varepsilon}_t$, where $\bm{\varepsilon}_t\sim \text{i.i.d.}\mathrm{N}(\bm{0},\bm{H})$ and $\bm{Z}(\cdot)$ is now a non-linear vector function. When performing our proposed update~\eqref{level update} using Gauss-Newton optimisation steps initialised at the prediction $\hat{\bm{x}}_{t|t-1}$, while ignoring (i.e.\ setting to zero) second-order derivatives of $\bm{Z}(\cdot)$, we obtain the iterated extended Kalman filter as a special case (for details, see \citealp[sec.~4.3]{lange2024bellman}). %If the observation noise $\bm{\varepsilon}_t$ is heavy tailed, however, the Bellman-filter update~\eqref{level update} suggests a `robustified' version of the Kalman filter and its (possibly iterated) extensions.
%, in which case the tail behaviour of $p(\bm{y}_t|\bm{x}_t)$ is accounted for.

\section{Bellman smoother equals Kalman smoother}

The Bellman smoother, derived in \citet[sec.~7]{lange2024bellman}, combines both forward- and backward dynamic-programming arguments. In the context of the linear Gaussian state equation~\eqref{state equation}, but a possibly non-linear and/or non-Gaussian observation equation~\eqref{observation equation}, the Bellman smoother turns out to be \emph{identical} to the classic Rauch-Tung-Striebel (RTS, \citeyear{rauch1965maximum}) smoother, which is often referred to as the Kalman smoother (e.g.\ \citealp[p.\ 91]{durbin2012time}). Assuming the data $\{\bm{y}_t\}_{t=1,\ldots,n}$ are fixed, the RTS smoother can be written as (e.g.\ \citealp[p.~154]{harvey1990forecasting} or \citealp[pp.~96-7]{durbin2012time}):
\begin{align}
\hat{\bm{x}}_{t|n} &=\hat{\bm{x}}_{t|t} +\bm{P}_{t|t} \bm{T}' \bm{P}^{-1}_{t+1|t} (\hat{\bm{x}}_{t+1|n}-\hat{\bm{x}}_{t+1|t}),
\label{RTS1}
\\
\bm{P}_{t|n}&=\bm{P}_{t|t}-\bm{P}_{t|t}\bm{T}' \bm{P}^{-1}_{t+1|t} ( \bm{P}_{t+1|t} - \bm{P}_{t+1|n} ) \bm{P}^{-1}_{t+1|t} \bm{T} \bm{P}_{t|t},
\label{RTS2}
\end{align}
which is a backward recursion for $t=1,\ldots,n$ that can be initialised at $t=n$ using the output of the filter. The only difference relative to the classic setting is that these expressions now apply in a possibly approximate context, where the forward recursion may be executed by running the Bellman (rather than the Kalman) filter. In the case of the linear Gaussian observation equation~\eqref{observation equation Kalman filter}, the Bellman filter/smoother combination collapses to the standard Kalman filter/smoother. 

\section{Static (hyper-)parameter estimation}

Apart from estimating the unobserved states, researchers also require a method to estimate the static (hyper-)parameter vector, denoted here as $\bm{\psi}$. The vector $\bm{\psi}$ contains (i) unknown shape parameters in the observation density $p(\bm{y}|\bm{x})$ and (ii) unknown static parameters in the state-transition equation~\eqref{state equation}, e.g.\ unknown parameters in $\bm{c}$, $\bm{T}$, $\bm{R}$ and $\bm{Q}$. 
%Shape parameters could actually be estimated by state augmentation, which means that these parameters are added
%by the Bellman filter augmentation, which 
An approximate parameter-estimation method based on the output of the Bellman filter, consisting of recursions~\eqref{level prediction} through \eqref{uncertainty update}, is provided in \citet[sec.~7]{lange2024bellman} as
\begin{equation}
\footnotesize 
\widehat{\bm{\psi}}:= \underset{\bm{\psi}}{\arg \max}
\sum_{t=1}^{n} \Bigg\{  \log p(\bm{y}_t|\hat{\bm{x}}_{t|t}) -\Big[\frac{1}{2} \log \frac{\det( \bm{P}_{t|t-1})}{\det( \bm{P}_{t|t})}+  \frac{1}{2}(\hat{\bm{x}}_{t|t}-\hat{\bm{x}}_{t|t-1})'\,\bm{P}^{-1}_{t|t-1}\,(\hat{\bm{x}}_{t|t}-\hat{\bm{x}}_{t|t-1})\Big] \Bigg\}.
\label{MLE}
\end{equation}
This estimator equals the standard maximum-likelihood estimator in the setting of the linear Gaussian observation equation~\eqref{observation equation Kalman filter}; otherwise, it can be viewed as maximising a second-order approximation of the standard prediction-error decomposition. Estimator~\eqref{MLE} can be implemented using standard numerical optimisation techniques, which is feasible as long as the dimension of $\bm{\psi}$ is not too large (e.g.\ not exceeding $20$). For a wide variety of models, \citet[sec.~8]{lange2024bellman} finds estimator~\eqref{MLE} to be roughly as accurate as (more computationally intensive) simulation-based methods, e.g.\ importance samplers and particle filters. 

\textbf{Interpretation of estimator~\eqref{MLE}.} The first term, $\log p(\bm{y}_t|\hat{\bm{x}}_{t|t})$, can be viewed as measuring the `goodness of fit' of the Bellman-filtered state $\hat{\bm{x}}_{t|t}$, relative to the observation $\bm{y}_t$. The term in square brackets, which appears with a minus sign in front of it, can be viewed as a penalty that penalizes deviations of $\hat{\bm{x}}_{t|t}$ from $\hat{\bm{x}}_{t|t-1}$ to prevent overfitting. 
%It resembles a `realised' version of the \citeauthor{kullback1951information} (KL, \citeyear{kullback1951information}) divergence, because it equals $[\log \phi(\bm{x};\hat{\bm{x}}_{t|t},\bm{P}_{t|t})-\log \phi(\bm{x};\hat{\bm{x}}_{t|t-1},\bm{P}_{t|t-1})]_{\bm{x}=\hat{\bm{x}}_{t|t}}$, where $\phi(\cdot;\bm{\mu},\bm{\Sigma})$ denotes the Gaussian density with mean $\bm{\mu}$ and covariance $\bm{\Sigma}$. To compute the (actual) KL divergence, the difference between both log-likelihood terms would have to be integrated; here, the difference is simply evaluated at $\bm{x}_{t|t}$, which means it can be viewed as a `realised' version of the KL divergence. 
The nonnegativity of this penalty follows from the nonnegativity of the quadratic expression along with the fact that $\det(\bm{P}_{t|t-1})\geq \det(\bm{P}_{t|t})$. The latter follows from uncertainty update~\eqref{uncertainty update}, which implies that $\bm{P}_{t|t-1}$ is weakly `larger', in a positive semi-definite sense, than $\bm{P}_{t|t}$.
%In turn, this is due to Fisher's information quantity $\bm{\mathcal{I}}(\hat{\bm{x}}_{t|t})$, which appears in the uncertainty update~\eqref{uncertainty update}, being positive semi-definite. 
In estimator~\eqref{MLE}, the trade-off between maximising the `fit' of the filtered states and minimising their `distance' from the predictions
%the realised KL divergence relative to the 
%the deviation from 
%relative to predicted states 
gives rise to a meaningful optimisation problem. 
%Resulting  parameter estimates are, in my experience, quite accurate.
%parameter estimates.

\section{Matrix-inversion lemmas}

The `matrix cookbook' (eqns.~156 and 158, which can be found \href{https://www.math.uwaterloo.ca/~hwolkowi/matrixcookbook.pdf}{here}) states two useful matrix-inversion lemmas for positive-definite matrices $\bm{A}$, $\bm{B}$ and an arbitrary (size-compatible) matrix~$\bm{C}$:
\begin{align}
 (\bm{A}+\bm{C}'\bm{B}^{-1}\bm{C})^{-1}&\bm{C}'\bm{B}^{-1}=\bm{A}^{-1}\bm{C}' (\bm{B}+\bm{C}\bm{A}^{-1}\bm{C}')^{-1}.
    \label{matrixlemma2}
    \\
    (\bm{A}+\bm{C}'\bm{B}^{-1}\bm{C})^{-1} &=\bm{A}^{-1}-\bm{A}^{-1}\bm{C}'(\bm{B}+\bm{C}\bm{A}^{-1}\bm{C}')^{-1}\bm{C}\bm{A}^{-1}.
    \label{matrixlemma1}
\end{align}
The second identity is a special case of the Woodbury matrix identity (e.g.\ \citealp{sherman1950adjustment} and \citealp{henderson1981deriving}). For the first identity, however, no proof is provided in the cookbook or the reference cited there. While the identity itself appears to be well-known (e.g.\ it is implied by equations 9 and 10 of \citealp{henderson1981deriving}), a direct reference is  hard to find. For clarity, I provide a joint proof of both identities.

For positive-definite (hence invertible) matrices $\bm{A}$ and $\bm{B}$ and an arbitrary (but size-compatible) matrix~$\bm{C}$, consider the following block matrix, $\bm{M}$, and its proposed inverse:
\begin{equation}
\footnotesize
   \bm{M}:= \left[\begin{array}{cc} \bm{A} & \bm{C}' \\ \bm{C} & -\bm{B} \end{array} \right],\quad
\label{inverse1}
\bm{M}^{-1}
 =  \left[\begin{array}{cc} (\bm{A}+\bm{C}'\bm{B}^{-1}\bm{C})^{-1} & (\bm{A}+\bm{C}'\bm{B}^{-1}\bm{C})^{-1}\bm{C}'\bm{B}^{-1} \\ (\bm{B}+\bm{C}\bm{A}^{-1}\bm{C}')^{-1}\bm{C}\bm{A}^{-1} & -(\bm{B}+\bm{C}\bm{A}^{-1}\bm{C}')^{-1}\end{array} \right].
\end{equation}
The correctness of the proposed inverse $\bm{M}^{-1}$ can be verified by computing $\bm{M}^{-1}\bm{M}$ to yield the identity matrix. Alternatively, $\bm{M}^{-1}$ can be expressed as
\begin{equation}
\small
\bm{M}^{-1}
 =  \left[\begin{array}{cc} \bm{A}^{-1}-\bm{A}^{-1}\bm{C}'(\bm{B}+\bm{C}\bm{A}^{-1}\bm{C}')^{-1}\bm{C}\bm{A}^{-1} & \bm{A}^{-1}\bm{C}'(\bm{B}+\bm{C}\bm{A}^{-1}\bm{C}')^{-1} \\ (\bm{B}+\bm{C}\bm{A}^{-1}\bm{C}')^{-1}\bm{C}\bm{A}^{-1} & -(\bm{B}+\bm{C}\bm{A}^{-1}\bm{C}')^{-1}\end{array} \right],
 \label{inverse2}
\end{equation}
which can again be verified by computing $\bm{M}^{-1}\bm{M}$ to yield the identity. All required matrix inverses in expressions~\eqref{inverse1}
and~\eqref{inverse2} exist, relying only on the (assumed) positive definiteness of $\bm{A}$ and $\bm{B}$. Hence $\bm{M}^{-1}$ exists. As $\bm{M}^{-1}$ is unique, the expressions in equations~\eqref{inverse1} and~\eqref{inverse2} must be equal. Comparing both top-right blocks yields equation~\eqref{matrixlemma2}, while comparing the top-left blocks yields equation~\eqref{matrixlemma1}.
%\footnote{Computing $\bm{M}\bm{M}^{-1}$ (rather than $\bm{M}^{-1}\bm{M}$), using either expression~\eqref{inverse1} or~\eqref{inverse2}, and setting the result equal to the identity matrix may yield further results.}

\section{Computational complexity}

The computational complexity of the Bellman filter and smoother is similar to that of the \citeauthor{kalman1960new} filter and smoother, enabling scalability to high-dimensional state spaces, e.g.\ $150$ dimensions in \citet[sec.~9]{lange2024bellman}. In such high-dimensional settings, it is advisable to implement the Bellman filter update~\eqref{level update} using a computationally efficient optimisation technique that avoids repeated large-matrix inversions, e.g.\ the Broyden–Fletcher–Goldfarb– Shanno (BFGS) algorithm. The Bellman filter's approach to `filtering by optimisation' avoids the curse of dimensionality faced by other methods (e.g.\ particle filters) that aim to approximate the mean. Intuitively, optimising a smooth strongly concave function in $150$ dimensions is easier than integrating it. \citet[sec.~9]{lange2024bellman} finds that, in such a high-dimensional setting, the Bellman filter can outperform the particle filter in terms of both speed and precision: the former by a factor ${\sim}10^3$ and the latter by a factor of ${\sim}2$, even when the particle filter employs very many particles (e.g.\ one million).

\section{Computer code}

\texttt{Matlab} code for Bellman filtering, smoothing and parameter estimation for several state-space models is available from the \emph{Journal of Econometrics} \href{https://doi.org/10.1016/j.jeconom.2023.105632}{website} or my \href{https://sites.google.com/view/rutgerjanlange/home}{homepage}.

\bibliographystyle{Chicago}

{\footnotesize {\ \setlength{\bibsep}{0pt plus 0.6ex}
\bibliography{references}
}}

\end{document}